\documentclass[aps,prc,12pt,onecolumn,nofootinbib,superscriptaddress,showpacs]{revtex4}
\usepackage{epsfig}
\usepackage{verbatim}
\usepackage{amssymb,amsmath}
\usepackage[usenames]{color}
\def\be{\begin{equation}} 
\def\ee{\end{equation}} 
\def\bea{\begin{eqnarray}}
\def\eea{\end{eqnarray}}
\def\nnb{\nonumber}

\def\half{{\textstyle\frac12}}

\def\etal{{\it et al}}

\def\N{{\scriptscriptstyle N}}
\def\V{{\scriptscriptstyle V}}
\def\S{{\scriptscriptstyle S}}
\def\mpi{m_\pi}

\def\mn{M_\N}
\def\fpi{f_\pi}

\def\ga{g_{\scriptscriptstyle A}}
\def\muv{\mu_\V}
\def\mus{\mu_\S}

\def\Tr{{\rm Tr}}

\def\slash#1{#1\llap/}

\def\Im{{\rm Im}}

\def\manchester{Theoretical Physics Group, 
School of Physics and Astronomy, \\
The University of Manchester,
Manchester, M13 9PL, UK}
\def\osaka{Department of Physics, 
Osaka University, Toyonaka, \\
Osaka 560-0043, Japan}

\begin{document}

\title{\Large\bf The $D$ coefficient in neutron beta decay \\
in effective field theory}

\author{Shung-ichi Ando}
\email{shung-ichi.ando@manchester.ac.uk}
\affiliation{\manchester}
\author{Judith A. McGovern}
\email{judith.mcgovern@manchester.ac.uk}
\affiliation{\manchester}
\author{Toru Sato}
\email{tsato@phys.sci.osaka-u.ac.jp}
\affiliation{\osaka}
\date{\today}
\begin{abstract}

In this paper we explore the time-reversal-odd triple-correlation coefficient in neutron beta decay, the so-called ``$D$~coefficient", 
using heavy-baryon effective field theory with photon degrees of freedom. We find that this framework allows us to reproduce the known results for the contribution which comes from final-state interactions, and also to discuss higher-order corrections.  In particular we are able to show that in the heavy-baryon limit all electromagnetic contributions vanish. By calculating the leading correction to the known result, we give a final expression which is accurate to better than 1\%. Hence we extend downwards the range over which the $D$ coefficient could be used to explore time-reversal violation from new physics.

\end{abstract}


\maketitle

\newpage

\section{Introduction}

In the expression for the polarised-neutron beta-decay rate, one of the terms which can be written is a triple scalar product of the neutron spin polarization and the momenta of the electron and neutrino; its coefficient is referred to as the ``$D$ coefficient"
~\cite{jtw-pr57}.  Such a term is time-reversal-odd and, in the absence of final-state interactions, a non-zero coefficient would indicate the presence of time-reversal violating interactions in the beta-decay vertex. 
 
An estimate of the $D$ coefficient from  CP-violation in the standard model 
turns out to be extremely tiny, $D_{\rm SM}\le 10^{-12}$~\cite{hk-prd97},
whereas a prediction from, e.g., the minimal supersymmetric
standard model (MSSM) is reported as 
$D_{\rm MSSM}\sim 10^{-7}$~\cite{dr-epjc03} and more exotic mechanisms such as leptoquarks could give even larger results~\cite{h-ppnp01}.
Experimental measurements of the $D$ coefficient may soon probe these values; recent papers 
give bounds of 
$D=[-0.6\pm1.2(\rm stat)\pm0.5(\rm syst)]\times 10^{-3}$ from the {\sc emi}T experiment \cite{letal-prc00} 
and 
$D=[-2.8\pm6.4(\rm stat)\pm3.0(\rm syst)]\times 10^{-4}$ from the Trine collaboration \cite{setal-plb04};
for a review of the future experimental prospects see, e.g., 
Ref.~\cite{ns-apnps05}. Increasingly precise measurements could open a new era for studies of physics beyond the standard model.
A complication is that there are small but non-zero corrections
from the electromagnetic final-state interaction~\cite{jtw-pr57}.  The correction 
at the one-photon loop order vanishes in the zero-recoil approximation, and (relaxing that approximation) has been estimated by Callan and Treiman as $D_{\rm FS}\simeq  10^{-5}$~\cite{ct-pr67}.\footnote{
For a historical review, see Ref.~\cite{t-arnps96}. This result has been confirmed by Refs~\cite{Brod70,Hol72,i-ptp78}}   If new physics which contributes at or below this magnitude is to be explored, it would be desirable to know this value to an accuracy of 1\% 
or less.  As previously established in many other processes, the ideal tool for such work is low-energy effective field theory (EFT)~\cite{erm-ppnp05}. 

In this work, we calculate the $D$ coefficient employing a pionless heavy-baryon EFT based on heavy-baryon chiral perturbation theory (HB$\chi$PT) with photon fields~\cite{w-pa79,jm-plb91,mm-npb99}.
HB$\chi$PT is a low-energy effective field theory of QCD
and has a systematic expansion scheme (counting rules) 
in terms of small external momenta, symmetry-breaking terms (proportional to the pion mass, $\mpi$),
and the number of loops. ``Small" 
is in relation to the scale $\Lambda_\chi\approx m_\rho$, which characterises the physics which is not explicitly included.  ``Heavy-baryon" indicates the approach in which the nucleon mass $\mn$ is also treated as a ``large" 
scale, $\mn\approx\Lambda_\chi$.  Hence the usual expansion parameter of HB$\chi$PT is roughly of order $\mpi/\mn$.  (For a recent review see Ref.~\cite{Ber08}.) However, because
the typical energy of neutron beta decay is so small
compared even to $m_\pi$, the pion itself can be integrated out and is not included as an explicit degree of freedom.  The counting rules for neutron decay in the pionless EFT have been worked out by Ando \etal~\cite{aetal-plb04}.
%the ordinary counting rules in HB$\chi$PT should be modified for the process~\cite{aetal-plb04}. 
We employ $\alpha/2\pi$ (which governs radiative corrections associated with photon loops), and $\bar{Q}/\mn$ (from the heavy-baryon expansion)
as our expansion parameters, where $\alpha$ is 
the fine structure constant 
and  $\bar{Q}$ is a typical momentum of neutron beta decay,
$\bar{Q}\sim m_n-m_p - m_e$ ($m_n,$ $m_p$ and $m_e$ are 
the neutron, proton, and electron mass, respectively).
These two expansion parameters are numerically almost the same,
$\alpha/2\pi\sim \bar{Q}/\mn\sim 10^{-3}$.

As we will show the Callan-Treiman result is reproduced in the EFT framework as the ${\cal O}(\alpha \bar{Q}/\mn)$ contribution,\footnote{Since, as we will see, the $D$ coefficient requires a loop integral with an imaginary part, contributions to it will be generically a factor of $2\pi$ larger than other terms at the same loop or $1/\mn$ order, which we will keep track of: hence ${\cal O}(\alpha \bar{Q}/\mn)$ rather than ${\cal O}(\alpha \bar{Q}/2\pi\mn)$.} but the counting suggests that ${\cal O}(\alpha^2/2\pi)$ contributions could be as large. Such contributions coming from repeated Coulomb final-state interactions have already been shown to vanish for zero recoil \cite{Brod70}, but radiative effects have not.  However we are able to show that in the limit $\mn\to \infty$ {\it all} electromagnetic contributions to the $D$ coefficient vanish, and so corrections to the Callan-Treiman result must be higher order.

In fact the counting rules discussed above, while valid for the first three orders, must be modified at the order at which the counterterms encode information about the pions which have been integrated out. In fact the leading higher-order correction to the $D$ coefficient, which we calculate, comes from the pseudoscalar form factor of the weak nucleon current, which is enhanced because its scale is governed not by the nucleon mass but by the pion mass \cite{BKM94}. This allows us to conclude that the residual error is of the order of 1 part in $10^{3}$, and certainly smaller than 1\%.

\section{The $D$ coefficient and effective Lagrangian}

The general expression for the differential neutron decay rate $d\Gamma$
is well known for the case wherein only the neutron is polarized:
\bea
\frac{d\Gamma}{dE_e d\Omega_{\hat{p}_e}d\Omega_{\hat{p}_\nu}}
&\simeq& \frac{(G_FV_{ud})^2}{(2\pi)^5}(1+3\ga^2) |\vec{p}_e|E_eE_\nu^2
\nnb \\ &&
\times \left(
1
+ a 
\frac{\vec{p}_e\cdot \vec{p}_\nu}{E_eE_\nu}
+ \hat{n}\cdot \left(
A \frac{\vec{p}_e}{E_e}
+ B \frac{\vec{p}_\nu}{E_\nu}
+ D \frac{\vec{p}_e\times\vec{p}_\nu}{E_eE_\nu}\right)
\right)\, ,
\label{eq;aABD}
\eea
where $G_F$ is the Fermi constant, $V_{ud}$ is a CKM matrix 
element, and $\ga$ is the axial-current coupling constant.
Here $E_e$ and $\vec{p}_e$ ($E_\nu$ and $\vec{p}_\nu$) are the 
electron (neutrino) energy and momentum,
$\hat{n}$ is the neutron spin polarization vector,
and $a$, $A$, $B$, $D$ are the correlation coefficients.
The standard lowest order 
expressions for the correlation coefficients are 
\bea
a = \frac{1-\ga^2}{1+3\ga^2}\, ,
\ \ \ 
A = \frac{-2\ga^2+2\ga}{1+3\ga^2}\, ,
\ \ \ 
B = \frac{2\ga^2+2\ga}{1+3\ga^2}\, ,
\ \ \
D = 0\, . \label{eq;aABD2}
\eea
We can reproduce these expressions for the coefficients in 
the leading order (LO) EFT calculation.
In addition, at NLO the ${\cal O}(\alpha/2\pi)$
radiative corrections and the ${\cal O}(\bar{Q}/\mn)$ recoil corrections
(including the weak magnetism term)
to the decay rate $\Gamma$ and 
correlation coefficients 
$a$, $A$, and $B$ 
have already been reported in Ref.~\cite{aetal-plb04}. 
A non-zero contribution to the $D$ coefficient,
due to the electromagnetic final-state interaction, 
has been obtained by Callan and Treiman 
from one-photon-loop contributions~\cite{ct-pr67} and
we will see that the same expression comes out at NNLO.

The standard chiral counting rules for 
HB$\chi$PT with and without photons
are discussed in Refs.~\cite{w-pa79,jm-plb91,mm-npb99}.  However as mentioned 
above, and discussed in more detail in Ref.~\cite{aetal-plb04}, the smallness of the energy released in neutron beta decay modifies the counting so that we have two parameters, $\bar{Q}/\mn\simeq \bar{Q}/\Lambda_\chi$ and $\alpha/2\pi$, which are is numerically of the same magnitude.  Thus we treat them as a single common expansion parameter in defining LO, NLO, etc.

The effective Lagrangian for the calculation of the $D$
coefficient in neutron beta decay reads~\cite{aetal-plb04}
\bea
{\cal L}_\beta =
{\cal L}_{e\nu\gamma}
+ {\cal L}_{NN\gamma}
+ {\cal L}_{NNe\nu}\, ,
\eea
where ${\cal L}_{e\nu\gamma}$ is the standard 
Lagrangian for 
electron, neutrino, and photon, 
${\cal L}_{NN\gamma}$ is the heavy-baryon Lagrangian for nucleon
interaction with a photon up to NLO
(the $1/\mn$ order), and ${\cal L}_{NNe\nu}$ is the $V$-$A$ interaction
Lagrangian between the nucleon and lepton currents
up to NLO:
\bea
{\cal L}_{e\nu\gamma} &=& 
- \frac14F^{\alpha\beta}F_{\alpha\beta}
- \frac{1}{2\xi_A} (\partial\cdot A)^2 
+ \bar{\psi}_e (i\gamma\cdot D -m_e)\psi_e
+ \bar{\psi}_\nu i\gamma\cdot\partial \psi_\nu\, , 
\\
{\cal L}_{NN\gamma} &=& N^\dagger iv\cdot DN 
\nnb \\ &&
+\frac{1}{2\mn}N^\dagger 
\Bigl(
(v\cdot D)^2-D^2 
-\frac i 2[S^\alpha,S^\beta]\bigl(
\mu_\V \tilde f_{\alpha\beta}^+
+ \half \mu_\S  \Tr[f_{\alpha\beta}^+]
\bigr) \Bigr) N \, ,
\\
{\cal L}_{NNe\nu} &=&  -\frac{G_FV_{ud}}{\sqrt2}
\bar{\psi}_e \gamma_\alpha(1-\gamma_5)\psi_\nu\Bigl(
N^\dagger \tau^+[v^\alpha-2\ga S^\alpha]N
   \nnb \\ && 
+\frac{1}{2\mn} N^\dagger\tau^+ \left(
i(v^\alpha v^\beta-g^{\alpha\beta}-2\ga v^\alpha S^\beta)
(\stackrel{\leftarrow}{\partial}-\stackrel{\to}{\partial})_\beta
-2i\muv[S^\alpha,S\cdot (\stackrel{\leftarrow}{\partial}
 +\stackrel{\to}{\partial})]
\right)N \Bigr) \, ,
\nnb \\
\eea
where $F_{\alpha\beta}=\partial_\alpha A_\beta - \partial_\beta A_\alpha$
and $D_\alpha$ is the covariant derivative of QED.
$\xi_A$ is the gauge parameter and we choose the Feynman gauge
$\xi_A=1$.
$v^\alpha$ is the velocity vector with the condition
$v^2=1$. 
In the rest frame of the neutron, $v^\mu=(1,\vec{0})$ and 
$2S^\mu=(0,\vec{\sigma})$.
Furthermore, the photon couples to the nucleon via the charge operator ${\cal Q}=\half(1+\tau_3)$, giving
$f_{\alpha\beta}^+ = 2{\cal Q} F_{\alpha\beta}$ and $\tilde f^+\equiv f^+-\half\Tr[f^+]$. $\muv$ and $\mus$ are 
isovector and isoscalar nucleon magnetic moments; $\muv = \mu_p-\mu_n = 4.706$.

The counting rules discussed above implicitly assume that higher-order terms in the Lagrangian are suppressed by successively more powers of $\Lambda_\chi\approx \mn$.  In reality however the break-down scale of the EFT is governed by the lightest degree of freedom which has been (at least conceptually) integrated out.  Here, this is the pion, which is substantially lighter than $\Lambda_\chi$.  The low-energy constants (LECs) of the pionless theory may therefore be governed by inverse powers of the pion mass, and hence larger than expected.

To the order we are working, the only LECs which enter are coupling constants such as $\ga$ and the magnetic moments, and these are fixed by their well-known experimental values.  At higher orders though, potentially $1/\mpi$-enhanced form-factor terms start to show up.  In fact at the next order there is only one such term, the induced pseudoscalar form factor of the nucleon. This is discussed in the final section.

\section{Amplitudes}

\begin{figure}
\begin{center}
\epsfig{file=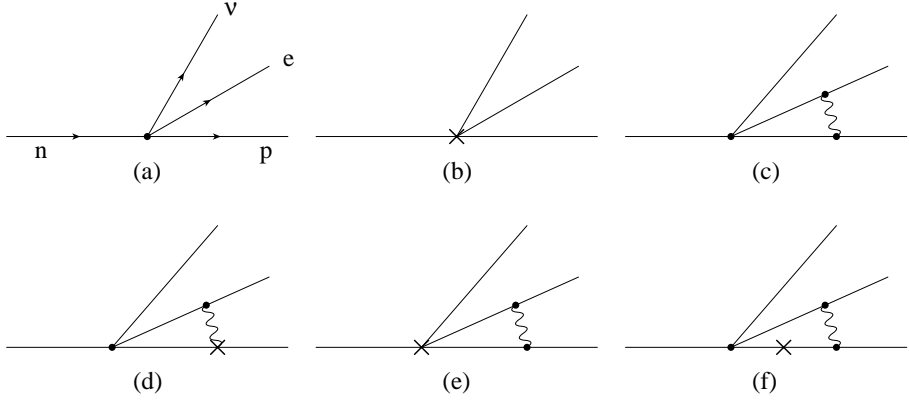,width=12cm}
\caption{Diagrams for the ${\cal O}(\alpha\bar{Q}/\mn)$
contributions to the $D$ coefficient. In all diagrams the LO nucleon vertices 
are denoted with a dot, and  NLO ones
(${\cal O}(\bar{Q}/\mn)$) with a cross. A wavy line denotes a photon.
Diagrams (a) and (b) are tree-level 
LO and NLO diagrams.
Diagram (c) is NLO (${\cal O}(\alpha2\pi)$) while
diagrams (d), (e), and (f) are 
NNLO as they include ${\cal O}(\bar{Q}/\mn)$ corrections.
}
\label{fig;beta-D}
\end{center}
\end{figure}
\begin{figure}[t]
\begin{center}
\epsfig{file=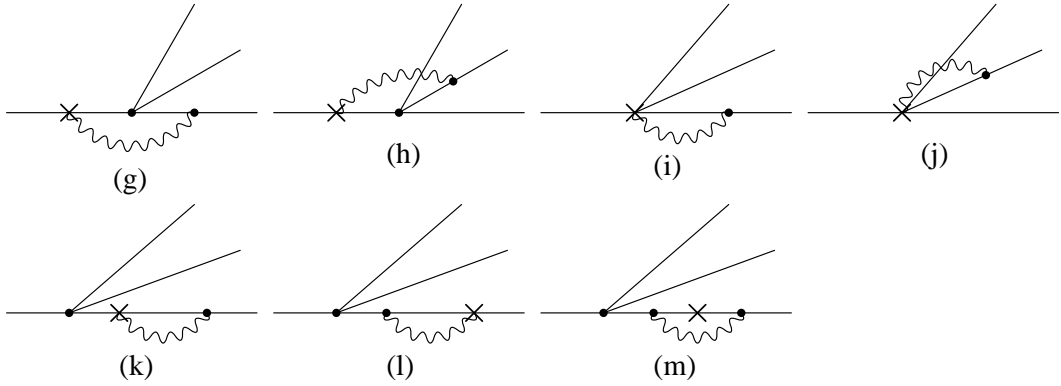,width=14cm}
\caption{
Diagrams for the ${\cal O}(\alpha\bar{Q}/\mn)$ 
amplitudes which do not contribute to the $D$ coefficient. 
See the caption of Fig.~\ref{fig;beta-D} for the details.
}
\label{fig;beta-D-1overmN}
\end{center}
\end{figure}
\begin{figure}[t]
\begin{center}
\epsfig{file=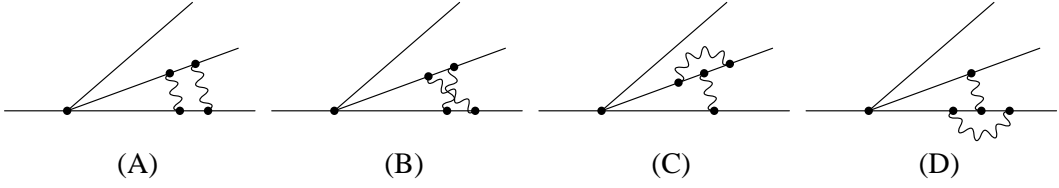,width=14cm}
\caption{
The non-trivial diagrams for the ${\cal O}(\alpha^2/2\pi)$ amplitudes. 
See the caption of Fig.~\ref{fig;beta-D} for details.
}
\label{fig;alpha2}
\end{center}
\end{figure}

The relevant Feynman diagrams for the NNLO calculation
are shown in Figs.~\ref{fig;beta-D},
\ref{fig;beta-D-1overmN}, and \ref{fig;alpha2}.
A contribution to the $D$ coefficient is obtained from 
the interference between the imaginary part of an amplitude 
from a loop diagram and the real amplitude of
a tree-level diagram~\cite{jtw-pr57,ct-pr67}.  
Fig.~\ref{fig;beta-D} contains the tree diagrams and those loop diagrams which have an imaginary part and which contribute to the $D$ coefficient; these together give the ${\cal O}(\alpha\bar{Q}/\mn)$ contribution.
The others do not contribute, in the case of Fig.~\ref{fig;beta-D-1overmN} because they have no imaginary part.  The diagrams of Fig.~\ref{fig;alpha2} (which with the leading tree graph would give ${\cal O}(\alpha^2/2\pi)$ contributions) will be considered in section~5.  There is no ${\cal O}\bigl((\bar{Q}/\mn)^2\bigr)$ contribution since this comes from tree amplitudes only.

We first consider the tree-level amplitudes.
The LO amplitude from the diagram (a) in Fig.~\ref{fig;beta-D} reads 
\bea
M_{(a)} &=& \frac{G_FV_{ud}}{\sqrt{2}}
\bar{u}_e(p_e)\gamma_\alpha(1-\gamma_5)v_\nu(p_\nu) 
\chi_p^\dagger (v^\alpha-2\ga S^\alpha)\chi_n 
\nnb \\ &\equiv&
 \frac{G_FV_{ud}}{\sqrt2}M_0\, ,
\eea
The ${\cal O}(\bar{Q}/\mn)$ tree-level amplitude from the diagram (b) 
in Fig.~\ref{fig;beta-D} reads
\bea
M_{(b)} &=& \frac{G_FV_{ud}}{\sqrt{2}}
\bar{u}_e(p_e)\gamma_\alpha(1-\gamma_5)v_\nu(p_\nu)  
\nnb \\ &&
\times\frac{1}{2\mn}\chi_p^\dagger \left(
(g^{\alpha\beta}-v^\alpha v^\beta+2\ga v^\alpha S^\beta)(p_p+p_n)_\beta
+2\muv [S^\alpha, S\cdot (p_p-p_n)]
\right) 
\chi_n\,,
\eea
where $p_p^\alpha$ and $p_n^\alpha$ are the {\it residual} four-momenta
for proton and neutron, respectively. 
We define them as 
$p_{n,p}^\alpha= P_{n,p}^\alpha-m_pv^\alpha$ 
where $P_n^\alpha$ and $P_p^\alpha$ 
are four-momenta for neutron and proton, respectively:
$P_n^2=m_n^2$ and $P_p^2=m_p^2$. 

We now consider the loop diagrams  
in Fig. \ref{fig;beta-D}.
The diagram (c) in Fig.~\ref{fig;beta-D} has 
already been calculated in Ref.~\cite{aetal-plb04}
giving
\bea
M_{(c)} &=& e^2\frac{G_FV_{ud}}{\sqrt{2}}\bigl(
M_0 F_0 +M_1 (-m_ef_2)
\bigr)\, ,
\eea
with
\bea
M_1 &=& \bar{u}_e(p_e) \slash{v} \gamma^\alpha(1-\gamma_5)v_\nu(p_\nu) 
\chi_p^\dagger (v_\alpha -2\ga S_\alpha)\chi_n\, ,
\label{eq;M1}
\\
F_0 &=& f_1 + 2E_e(f_2-f_0)\, ,
\label{eq;F0}
\eea
where $f_0$, $f_1$, and $f_2$ are loop functions
whose definitions are given in Appendix A.

From the diagrams (d), (e), and (f) in Fig.~\ref{fig;beta-D},
we have
\bea
M_{(d)} &=&
-\frac{G_FV_{ud}}{\sqrt{2}} \frac{e^2}{2\mn}
\int_l \frac{\bar{u}_e\gamma^\mu(\slash{p}_e-\slash{l}+m_e)
\gamma^\alpha(1-\gamma_5)v_\nu}{v\cdot l\, l^2\,(l^2-2p_e\cdot l)}
\nnb \\ && 
\times\chi_p^\dagger \left(
(g_{\mu\nu}-v_\mu v_\nu)(l+2p_p)^\nu
-2\mu_p[S_\mu,S\cdot l]\right)
(v_\alpha -2\ga  S_\alpha)\chi_n \, ,
\\
M_{(e)} &=& 
-\frac{G_FV_{ud}}{\sqrt{2}}\frac{e^2}{2\mn}
\int_l \frac{\bar{u}_e \slash{v}(\slash{p}_e-\slash{l}+m_e)
\gamma^\alpha(1-\gamma_5)v_\nu}{v\cdot l\,l^2\,(l^2-2p_e\cdot l)}
\nnb \\ && 
\times\chi_p^\dagger 
(g_{\alpha\beta}-v_\alpha v_\beta
+ 2\ga  v_\alpha S_\beta+2\muv [S_\alpha,S_\beta])(l+p_p+p_n)^\beta
 \chi_n \, ,
\nnb \\
\\
M_{(f)} &=& 
-\frac{G_FV_{ud}}{\sqrt{2}}\frac{e^2}{2\mn}
\int_l \frac{\bar{u}_e\slash{v}(\slash{p}_e-\slash{l}+m_e)
\gamma^\alpha(1-\gamma_5)v_\nu}{(v\cdot l)^2 \,l^2\, (l^2-2p_e\cdot l)}
(v\cdot l^2-l^2 -2p_p\cdot l)
\nnb \\ && 
\times\chi_p^\dagger (v_\alpha-2\ga S_\alpha)\chi_n \, ,
\eea
with
\bea
\int_l\equiv \frac{1}{i}\int\frac{d^4l}{(2\pi)^4}\, ,
\nnb 
\eea
where we have used the fact that the nucleon kinetic energy is ${\cal O}(\bar Q^2/\mn)$ to set $v\cdot p_p\simeq 0$ in the nucleon propagator.

\section{The $D$ coefficient from Heavy-Baryon EFT}

In calculating the decay rate, which is proportional to the modulus of the amplitude squared, we are taking the product of one diagram and the time-reversed version of another, e.g.\ $M_{(a)} M^*_{(d)}$.  Considering the same pair in the opposite order, $M_{(d)} M^*_{(a)}$, it is obvious that the time-reversal odd term $i \hat{n}\cdot (\vec{p}_e\times \vec{p}_\nu)$ will have the opposite sign and a complex-conjugated coefficient %in the second 
compared with the original order.  Thus when we sum the two, we can only get a non-zero result if the coefficient has an imaginary part, and this must come from a loop integral with a physical intermediate state. 

From the interference between the LO amplitude from the 
diagram (a) and the imaginary part of the leading loop diagram
(c), we have as the potential NLO contribution to the $D$ coefficient
\bea
D\propto \sum_{\rm spin}^D \left(
M_{(c)} M_{(a)}^* + M_{(a)} M^*_{(c)}
\right) = 0\, ,
\eea
where the superscript $D$ on the summation denotes that
we sum over the spins of the the electron, neutrino, and proton but not that of the neutron, and keep only terms contributing to the $D$ coefficient, those
proportional to 
$\hat{n}\cdot (\vec{p}_e\times \vec{p}_\nu)$ where $\hat{n}=\chi_n^\dagger \vec{\sigma}\chi_n$.  (The corresponding 4-vector spin polarisation $N^\mu$ is given by $N^\mu=\chi_n^\dagger 2S^\mu\chi_n$, and $i\epsilon^{\mu\nu\alpha\beta}
v_\mu N_\nu p_{e\alpha}p_{\nu\beta}=\hat{n}\cdot (\vec{p}_e\times \vec{p}_\nu)$.)  This NLO term vanishes because there is no such term in ${\sum_{\rm spin}} M_0^*M_{0,1}$,
and not through details of the loop integrals $F_0$ and $f_1$. Thus the $D$ coefficient has no contribution up to NLO.

We now consider ${\cal O}(\alpha \bar{Q}/\mn)$ interference terms.
From the amplitudes from the NLO (${\cal O}(\bar{Q}/\mn$)) diagram (b) and leading
loop diagram (c),
we have 
\bea
&&\sum_{\rm spin}^D\left(
M_{(c)}  M_{(b)}^* 
+ M_{(b)}  M^*_{(c)}\right) =
-4(G_FV_{ud})^2 
\frac{e^2}{2\mn} \epsilon^{\mu\nu\alpha\beta}
v_\mu N_\nu p_{e\alpha}p_{\nu\beta} 
\nnb \\ &&\hspace*{2cm} 
\times\left\{
2\Im F_0 \left( 
E_e\left( 
-\ga \muv
+\ga 
+\ga^2
-\muv
\right) 
+E_\nu\left( 
-\ga \muv
+\ga
-\ga^2
+\muv
\right) \right)\hspace*{1.3cm} \right.
\nnb \\ &&\hspace*{2.5cm} \left. 
+2\Im f_2 m_e^2\left(
\ga\muv
-\ga
-\ga^2
+\muv\right) \right\} \,.
\label{eq;D1}
\eea
From the LO amplitude $M_{(a)}$ and the NNLO 
(${\cal O}(\alpha\bar{Q}/\mn)$ amplitudes $M_{(d,e,f)}$, we have
\bea 
&&\sum_{\rm spin}^D 
\left(M_{(d)}M_{(a)}^* 
+ M_{(a)} M^*_{(d)}\right)  =
-4(G_FV_{ud})^2 \frac{e^2}{2\mn}\epsilon^{\mu\nu\alpha\beta}
v_\mu N_\nu p_{e\alpha}p_{\nu \beta}
\nnb \\ &&\hspace*{2cm} 
\times(1-\ga)\left(
-8\ga E_\nu \Im f_1
+2\mu_p (1+3\ga)\left(
2\Im f_4 +E_e \Im f_5 +m_e^2 \Im f_6 \right)
\right) \, ,
\label{eq;Dd}
\\ 
&&\sum_{\rm spin}^D \left(
M_{(e)}M_{(a)}^* + M_{(a)} M^*_{(e)}
\right) =
-4(G_FV_{ud})^2 \frac{e^2}{2\mn}\epsilon^{\mu\nu\alpha\beta}
v_\mu N_\nu p_{e\alpha}p_{\nu \beta} 
\nnb \\ && \hspace*{2cm} \left\{
\times 2\Im F_0 \left(
E_e(\ga\muv-\ga-\ga^2 +\muv)
+E_\nu(\ga\muv-\ga+\ga^2 -\muv)\right)
\right.\nnb \\ &&\hspace*{2.5cm} \left.
+2(\ga\muv-\ga-\ga^2+\muv)\left(
-E_e\Im f_5
+(-m_e^2+2E_e^2)(\Im f_2-\Im f_6)\right)
\right.\nnb \\ && \hspace*{2.5cm} \left.
+4(\muv-\ga)\Im f_4 \right\}\, ,
\label{eq;De}
\\ 
&&\sum_{\rm spin}^D \left(
M_{(f)}M_{(a)}^* + M_{(a)} M^*_{(f)}
\right) 
= -4(G_FV_{ud})^2\frac{e^2}{2\mn} \epsilon^{\mu\nu\alpha\beta}
v_\mu N_\nu p_{e\alpha}p_{\nu \beta}
\left(
-8\ga(1-\ga) E_\nu \Im g_4
\right)\,,
\nnb \\
\label{eq;Df}
\eea
where explicit expressions for the loop functions $f_i$ and $g_4$
are given in the Appendix.  (Throughout, we have used unit normalization for the heavy-baryon nucleon spinors).

To obtain the $D$ coefficient, we have to divide out the common factor $4 (G_FV_{ud})^2 E_eE_\nu (1+3\ga^2)$ which also appears in the angle-independent term in the leading matrix element squared,
$\sum_{\rm spin}M^*_{(a)} M_{(a)}$.

We thus have the final expression for the ${\cal O}(\alpha\bar{Q}/\mn)$ contribution to the $D$ coefficient
for neutron beta decay:
\bea
D_{\rm CT} &=&
\frac{1}{1+3\ga^2}\frac{\alpha E_e}{4\mn}\frac1\beta\Bigl(
(1+3\ga)\left((\muv-\ga)-3\mu_p(1-\ga)\right)
\nnb \\ &&\hspace*{2.5cm}
+\frac{m_e^2}{E_e^2}\left(
(3+\ga)(\muv-\ga) + 3\mu_p (1-\ga)(1+3\ga)\right)
\Bigr)\,.
\label{eq;D}
\eea
where $\beta\equiv |\vec{p}_e|/E_e$.
This is exactly the result obtained by Callan and Treiman~\cite{ct-pr67}.

\vskip 3mm \noindent
\section{Terms higher-order in $\alpha$}

Since, as discussed above, the two scales 
$\alpha/2\pi$ and $\bar{Q}/\mn$ are of the same size, 
${\cal O}(\alpha^2/2\pi)$ terms, if they exist, 
could be as large as the well-known 
${\cal O}(\alpha\bar{Q}/\mn)$ 
contribution which we reproduced in the previous section.  
We therefore need to check these too.  
The relevant diagrams are shown in Fig.~\ref{fig;alpha2}, 
and their amplitudes are as follows:
\bea
M_{(A)}&\!\propto\!&
\int_{l}\int_{k}
\frac{
\bar{u}_e(p_e)\slash{v}\,
(\slash{p}_e+\slash{k}+m_e)\,\slash{v}\, (\slash{p}_e+\slash{l}+m_e)\,
\gamma_\alpha(1-\gamma_5)v_\nu(p_\nu)} {
v\cdot l\, v\cdot k \,k^2 \,(k-l)^2 \,(l^2-2p_e\cdot l) \, (k^2-2p_e\cdot k)}
\chi_p^\dagger (v^\alpha-2\ga S^\alpha)\chi_n  \,,
\nonumber\\
M_{(B)}&\!\propto\!&
\int_{l}\int_{k}\frac{\bar{u}_e(p_e)\slash{v}\,
(\slash{p}_e+\slash{k}+m_e)\,\slash{v} \,(\slash{p}_e+\slash{l}+m_e)\,
\gamma_\alpha(1-\gamma_5)v_\nu(p_\nu)}{v\cdot l\, 
v\cdot (l-k) \,k^2\, (k-l)^2\, (l^2-2p_e\cdot l) \,(k^2-2p_e\cdot k) }
\chi_p^\dagger (v^\alpha-2\ga S^\alpha)\chi_n \,,
\nonumber\\
M_{(C)}&\!\propto\!& 
\int_{l}\int_{k}\frac{\bar{u}_e(p_e)\gamma^\sigma\,
(\slash{p}_e+\slash{k}+m_e)\,\slash{v} \,
(\slash{p}_e+\slash{l}+\slash{k}+m_e)\,
\gamma_\sigma\,(\slash{p}_e+\slash{l}+m_e)\,
\gamma_\alpha(1-\gamma_5)v_\nu(p_\nu)}{
v\cdot l \,l^2\, k^2\, (l^2-2p_e\cdot l) \, 
(k^2-2p_e\cdot k)\, ((l+k)^2-2p_e\cdot (l+k))}
\nnb \\ && \times
\chi_p^\dagger (v^\alpha-2\ga S^\alpha)\chi_n\,,
\nonumber\\
M_{(D)}&\!\propto\!&
\int_{l}\int_{k}\frac{\bar{u}_e(p_e)\slash{v} \,
(\slash{p}_e+\slash{l}+m_e)\,
\gamma_\alpha(1-\gamma_5)v_\nu(p_\nu)}{v\cdot l\, 
v\cdot (l+k)\,v\cdot k\,l^2\, k^2\, (l^2-2p_e\cdot l)}
\chi_p^\dagger (v^\alpha-2\ga S^\alpha)\chi_n\,.
\eea
Clearly the structure of the amplitude $M_{(D)}$ is 
the same as the leading one-photon-loop diagram $M_{(c)}$, 
though the integral is different.  
So interference with $M_0$ cannot give a contribution to 
the $D$ coefficient. 
The other three cases are more complicated 
(though it is worth noticing that the tensor structures of 
$M_{(A)}$ and $M_{(B)}$ are the same; again only the integrals differ).  
However explicit calculation---without the need actually 
to calculate any scalar two-loop integrals---again shows 
a vanishing contribution to the $D$ coefficient.
(The calculations are most efficiently done using e.g.~the package ``Tracer" 
on Mathematica~\cite{Tracer}; 
we do not give details of the reduction to scalar 
integrals as in the Appendix because they are not required.)

Having shown through explicit calculation that 
there are no two-photon-loop contributions to the $D$ coefficient 
in the heavy-baryon limit, it is interesting to consider
if this can be generalized.  
The crucial features which make it possible are that in this limit, 
as can be seen in the expressions for the diagrams (c) and (A-D), 
the amplitude factorizes into a hadronic and leptonic part. 
Since the leading-order photon coupling to the proton is spin-independent, 
no matter how complicated the photon exchanges and dressings are,
the hadronic tensor is unchanged.  
Furthermore, since the heavy-baryon propagator does not depend 
on the three-momentum of the proton, the loop integrals can only 
pick up factors of the electron momentum and the velocity vector, 
limiting the complexity of the structures which can  appear 
in the leptonic tensor.  
Thus it is plausible that many-photon effects do not in fact 
generate new structures.   
Recall that the vanishing of the $D$ coefficient
in the one- and 
two-photon-loop case is due to the tensor structure, and not 
to details of the integrals involved.

In fact it was shown long ago that one class of corrections 
vanish to all order in $\alpha$, namely the repeated exchange 
of Coulomb photons~\cite{Brod70}.  In the heavy-baryon limit 
these only introduce a phase shift in the final-state wavefunction, 
and cannot induce a $D$ coefficient.
However in the relativistic theory crossed photons, or multiple 
overlapping dressings of either fermion line, 
cannot be accounted for so easily as the interaction kernel 
becomes arbitrarily complicated.  In the heavy-baryon formalism 
however the complexity is much reduced, as indicated above.  
As we will demonstrate in what follows, we can show that no 
corrections survive at any order.

The amplitude resulting from the exchange of any number of 
(non-magnetic) photons (crossed or not) between the electron and proton, 
and from photon-loop dressing of one or more photon-electron 
or photon-proton vertices, has the same general form 
as the single photon diagram (c), albeit with multiple integrals 
and many insertions along the electron line. 
The basic structure can be written schematically as follows:
\be
M_X\propto\left(
\int_{l_1}\ldots \int_{l_n}\frac{\bar{u}_e(p_e)  \hat O(\{l_i\})
\gamma_\alpha(1-\gamma_5)v_\nu(p_\nu)}{v\cdot l'_i
\ldots {l'_j}^2\ldots ({l'_k}^2-2p_e\cdot l'_k)\ldots  }\right)
\chi_p^\dagger (v^\alpha-2\ga S^\alpha)\chi_n \label{one-loop}\,,
\ee
where the $l'_i$ are linear combinations of the loop momenta $l_i$, 
and $\hat O(\{l_i\})$ is composed only of the building blocks 
$m_e$, $\{\slash{l}_i\}$, $\slash{p}_e$, $\slash{v}$ and 
$\gamma_\sigma\ldots \gamma^\sigma$ (the last being for 
a photon loop dressing one or more vertices, and with 
$\ldots$ standing for more of the same).

When the loop integrals are performed, 
all loop momenta $l_i^\alpha$ in the numerator  
either become $p_e^\alpha$ or $v^\alpha$, 
or a pair gives $l_i^\alpha l_j^\beta\to g^{\alpha\beta}$.  
Thus after integration, 
the whole structure represented by 
$\hat O(\{l_i\})$ can only give a small number of terms, namely 
$I$, $\slash{p}_e$, $\slash{v}$ and  $[\slash{p}_e,\slash{v}]$, 
and the same multiplied by $\gamma_5$.  
Furthermore since  
$\gamma_5\gamma_\alpha(1-\gamma_5)=\gamma_\alpha(1-\gamma_5)$, 
the structures with $\gamma_5$ in them are redundant.  
Hence the bottom line is 
\be
M_X=\sum_{n=1}^4 I^X_n(m_e,v\cdot p_e)\bar{u}_e(p_e)
\hat O_n\gamma_\alpha(1-\gamma_5)v_\nu(p_\nu) \chi_p^\dagger [v^\alpha-2\ga S^\alpha]\chi_n \,,
\ee
where the integrals  $I^X_n$ will depend on the particular 
graph we are considering, and the operators $\hat O_n$ are 
the four listed above.

When we calculate the decay rate from the amplitude 
we need expressions like $M_X^\dagger M_Y$,
with a sum over the spin of the proton, electron and neutrino.  
Using completeness relations such as 
$\sum_s u_e(p_e,s)\bar{u}_e(p_e,s)=\slash{p}_e+m_e$, we end up with
terms like the following:
\bea
M_X^\dagger M_Y&=&\sum_{m,n=1}^4 (I^X_m)^* I^Y_n\,
\Tr[\gamma_\alpha(1-\gamma_5)\bar{O}_m(\slash{p}_e+m_e)
\hat O_n\gamma_\beta (1-\gamma_5)\slash{p}_\nu]\times\nonumber\\
&&\qquad \qquad
\chi_n^\dagger(v^\alpha-2\ga S^\alpha)(v^\beta-2\ga S^\beta)\chi_n \,,
\eea
where $\bar{O}_m=\gamma_0 \hat O_m^\dagger \gamma_0 =\pm \hat O_m $.  
Finally, we note that $\bar{O}_m(\slash{p}_e+m_e)\hat O_n$ is itself 
just a combination of the $\hat O_m$, giving just four structures 
to be calculated:
\be
\Tr[\gamma_\alpha(1-\gamma_5)\hat O_n\gamma_\beta(1-\gamma_5)\slash{p}_\nu]\;
\chi_n^\dagger(v^\alpha-2\ga S^\alpha)(v^\beta-2\ga S^\beta)\chi_n \,.
\ee
Only $\hat O_n=\slash{p}_e$ and $\slash{v}$ give non-vanishing results, 
and they do not generate the structure 
$ \epsilon_{\alpha\beta\sigma\tau} p_e^{\alpha}p_\nu^{\beta}v^{\sigma}
\chi_n^\dagger S^{\tau}\chi_n $.  
So in the heavy-baryon limit, there are no contributions to 
the $D$ coefficient at any order in $\alpha$.
Corrections to the leading $\alpha\bar{Q}/\mn$ result will be 
${\cal O}(\alpha^2 \bar{Q}/2\pi\mn)$ and ${\cal O}(\alpha\bar{Q}^2/\mn^2)$ ---presumably at least another factor of $10^{-3}$ down.\footnote{The referee of this paper has drawn our attention to the paper by
Gross \cite{Gro82} on two particles of unequal mass interacting via
relativistic exchange of some boson, and in particular to the effective one-body Dirac equation obeyed by the light particle when the heavy particle's mass is taken to infinity.   It is possible that
the all-orders vanishing of electromagnetic contributions discussed here may also be explicable within that framework.}

\section{Higher-order corrections and conclusions}

In the preceding sections, we have shown that the Callan-Treiman result for the $D$ coefficient in neutron $\beta$ decay is the leading non-vanishing contribution in a heavy-baryon EFT in which the expansion parameters, $\alpha/2\pi$ and $\bar{Q}/\mn$, are both of the order of $10^{-3}$. The non-zero contribution to the $D$ coefficient appears because of the electromagnetic final-state interaction at ${\cal O}(\alpha\bar{Q}/\mn)$, 
whereas there are no contributions from the ${\cal O}(\alpha^2/2\pi)$ and ${\cal O}\bigl((\bar{Q}/\mn)^2\bigr)$ terms. The $D$ coefficient is NNLO with respect to the decay rate, and hence very small ($\sim 10^{-5}$).  Nonetheless there are hopes that this might be experimentally accessible in the not-too-distant future. The interest in such experiments would, of course, be to detect deviations which might indicate new physics, and to interpret such a result the accuracy of the Callan-Treiman prediction must be known.  At first glance our results suggest that the next correction would introduce a relative error of order $10^{-3}$, which should be small enough to be unimportant for many years to come; in particular this would be small enough to allow detection (if sufficiently sensitive experiments could be carried out) of the prediction from the MSSM of a contribution to the $D$ coefficient of the order of $10^{-7}$~\cite{dr-epjc03}.

However at the next order (N$^3$LO) there will be contributions from the third-order Lagrangian, and these will not only be the $1/\mn^2$ terms required by Lorentz invariance, but will also include new structures whose scale is governed by the lightest degrees of freedom which have been integrated out---in this case the pion.  Potentially therefore the corrections to the ${\cal O}(\alpha\bar{Q}/\mn)$ result are ${\cal O}(\alpha\bar{Q}^2/\mpi^2)$, giving a relative error of perhaps 5\%.

Looking at the diagrams of Fig.~1, such insertions from the third-order Lagrangian could in principle replace any of the crosses, though that in Fig.~1f could only give a proton mass shift. However because we need to take the imaginary part of the loop graphs, the nucleon is on-shell everywhere and most of the vertex corrections are just form-factor corrections to the structures we have already considered---e.g.\ $\mu_p\to\mu_p(1+\langle r_M^2\rangle q^2/6$)---which are therefore two orders down.\footnote{If we work in HB$\chi$PT with explicit pions, these form factors arise partly from pion loops.  The leading term which is  ${\cal O}(\bar{Q}\mpi/(4\pi \fpi^2)$  simply renormalises the isovector magnetic moment and the next term in an expansion in the photon momentum transfer gives a term of relative size $\bar{Q}^2/\mpi^2$.}  Only one potentially-enhanced third-order effect remains, and that is where the cross in Fig.~1b and Fig.~1e represents the pseudoscalar form factor of the nucleon, the leading  contribution to which is given by the exchange of a pion between the nucleon and leptons.

This pseudoscalar vertex is easily calculated from HB$\chi$PT to be as follows (where   $(q^2-\mpi^2)$ in the pion propagator has been replaced with $-\mpi^2$ since the momentum  transfer is very much less than $\mpi$):
\bea
-i M_{\rm PS}=i\frac{G_FV_{ud}}{\sqrt2}
\frac{2g_A }{m_\pi^2} q_\alpha S\cdot q\gamma^\alpha(1-\gamma_5) 
\, ,
\eea
There are two contributions to the $D$ coefficient, arising from the interference of the amplitudes (1a) and (1e), and of (1b) and (1c) (with the cross in (1b) and (1e) now representing the new vertex).  These give a final result of
\bea
D_\pi &=& - \frac{1}{1+3g_A^2} \frac{e^2m_e^2}{m_\pi^2} g_A^2\Bigl(
-4(E_e+E_\nu) \Im f_2 + 2\Im f_5
+ 2E_e\Im f_6 \Bigr)
\nnb \\ &=&
- \frac{1}{1+3g_A^2} \frac{\alpha m_e^2}{m_\pi^2}\frac{g_A^2}{\beta}
\frac{2(E_e+E_\nu)}{E_e}\,.
\label{eq;D-pi}\eea
With the approximation $E_e+E_\nu=m_n-m_p$ this has the value $-5.88(p_e^{\rm max}/p_e)\times 10^{-8}$
compared to the leading result of $(0.228(p_e^{\rm max}/p_e)+1.083(p_e/p_e^{\rm max}))\times 10^{-5}$---a correction of between $-0.5$\% and $-2.5$\%.\footnote{We have used the current PDG value of $\ga=1.2695\pm0.0029$.\cite{pdg}  The error on $\ga$ induces a 0.1\% 
error in $D$ at $p_e=p_e^{\rm max}$.}

With the only $(1/\mpi)$-enhanced N${}^3$LO term explicitly calculated, we can now say with confidence that our final result for the $D$ coefficient, 
\bea
D=D_{\rm CT}+D_\pi
\eea
where $D_{\rm CT}$ and $D_\pi$ are given by Eqs~(\ref{eq;D}) and (\ref{eq;D-pi}), should be accurate to better than 1\%,
with expected corrections being ${\cal O}(10^{-3})$.

\vskip 3mm \noindent
\begin{acknowledgments}

SA would like to thank 
J Behr and H~M Shimizu for discussions,
H-W Fearing and M Igarashi for communications. JAMcG would like to thank M Birse for discussions and for reading the manuscript and T Cohen for discussions.
The work of SA and JAMcG is supported by STFC grant number PP/F000448/1.
The work of TS is supported by the Japan Society for the Promotion of Science,
Grant-in-Aid for Scientific Research(c) 20540270.
\end{acknowledgments}

\appendix*{\center \bf Appendix: Loop functions}

The loop functions for the $D$-calculation in HB$\chi$PT are defined as
\bea
\int_l \frac{1}{(v\cdot l+i\eta)(l^2+i\eta)(l^2-2p_e\cdot l+i\eta)} 
&=& f_0\, ,
\eea
where
\bea
\int_l \equiv \frac{\mu^{4-d}}{i}\int \frac{d^dl}{(2\pi)^d}\, ,
\nnb
\eea
and $D$-is the space-time dimensions, $d=4-2\epsilon$.
Furthermore,
\bea
\int_l \frac{l^\mu}{v\cdot l l^2(l^2-2p_e\cdot l)} &=& 
v^\mu f_1 + p_e^\mu f_2 \, ,
\\
\int_l \frac{l^\mu l^\nu}{v\cdot l l^2(l^2-2p_e\cdot l)} &=& 
v^\mu v^\nu f_3 + g^{\mu\nu}f_4 
+ (v^\mu p_e^\nu + p_e^\mu v^\nu) f_5
+ p_e^\mu p_e^\nu f_6 \, ,
\\
\int_l \frac{l^\mu l^\nu}{(v\cdot l)^2l^2(l^2-2p_e\cdot l)}
&=& v^\mu v^\nu g_3 + g^{\mu\nu}g_4
+(v^\mu p_e^\nu +p_e^\mu v^\nu)g_5
+p_e^\mu p_e^\nu g_6\, .
\eea
where we have suppressed the $i\eta$ terms in
the propagators in those expressions.
To calculate the contribution to the $D$-term, we
need imaginary part of the six loop functions in the followings,
and thus have  
\bea &&
\Im  f_1 = -\Im g_4 = \frac{1}{8\pi}\frac{1}{\beta}\, ,
\ \ 
\Im f_2 = - \frac{1}{8\pi}\frac{1}{\beta E_e}\, ,
\nnb \\ && 
\Im f_4 = \frac{1}{16\pi}\beta E_e\, ,
\ \ 
\Im f_5 = \frac{1}{16\pi} \frac{1}{\beta}\, , 
\ \ 
\Im f_6 = - \frac{1}{16\pi}\frac{1}{\beta E_e}\, .
\eea

\end{document}